\documentclass[a4paper,11pt]{article}
\usepackage{pos}
\usepackage{natbib}
\setcitestyle{numbers,sort&compress}
\usepackage{xspace}
\usepackage{url}
\usepackage{hyperref}
\usepackage{aas_macros}
\usepackage{booktabs} 
\usepackage{orcidlink}
\usepackage[flushleft]{threeparttable}
 
\newcommand {\nicer}{{NICER}\xspace} 
\newcommand {\ixpe}{{IXPE}\xspace}
\newcommand {\nustar}{{NuSTAR}\xspace}
\newcommand {\hxmt}{{Insight-HXMT}\xspace}
\newcommand {\sco}{\mbox{{Sco~X-1}}\xspace} 
\newcommand\degr{\mbox{$^\circ$}}

\newcounter{daggerfootnote}
\newcommand*{\daggerfootnote}[1]{%
    \setcounter{daggerfootnote}{\value{footnote}}%
    \renewcommand*{\thefootnote}{}%
    \footnote[2]{#1}%
    \setcounter{footnote}{\value{daggerfootnote}}%
    \renewcommand*{\thefootnote}{\arabic{footnote}}%
    }

\title{Sco X-1 as seen by IXPE}

\author*[a,b,\dagger]{Fabio La Monaca$^{\orcidlink{0000-0001-8916-4156}}$ \daggerfootnote{\hspace{-0.15cm}$^\dagger$On behalf of the IXPE Science Team: {\href{https://ixpe.msfc.nasa.gov/partners_sci_team.html}{https://ixpe.msfc.nasa.gov/partners\_sci\_team.html}}}}

\affiliation[a]{INAF - Istituto di Astrofisica e Planetologia Spaziali\\
  Via del Fosso del Cavaliere~100, 00133 Rome, Italy}

\affiliation[b]{Dipartimento di Fisica, Universit\`{a} degli Studi di Roma ``Tor Vergata''\\
Via della Ricerca Scientifica 1, 00133 Rome, Italy}

\emailAdd{fabio.lamonaca@inaf.it}

\abstract{The X-ray polarization of \sco was measured for the first time with very high significance by the Imaging X-ray Polarimeter Explorer (IXPE). A polarization degree of $1.0\% \pm 0.2\%$ at a PA of $8\degr \pm 6\degr$ at 90\%~confidence level (CL) in the 2--8\,keV energy band is obtained, while the source was in its soft state with short flaring periods. The source state was determined by a strictly simultaneous X-ray observation campaign jointly with \ixpe, which involved \nicer, \nustar, and \hxmt, allowing for a broad-band spectrum characterization and study of quasiperiodic oscillations. The spectropolarimetric analysis yielded a polarization of $<3.2$\% for the accretion disk and a polarization of 1.3\% $\pm$ 0.4\% for the hard Comptonized component. A constraint on the polarization of the reflection component, modeled using \texttt{relxillNS}, is obtained. All the results about the polarization degree match the theoretical expectations, while the polarization angle of $8\degr \pm 6 \degr$ at 90\%\,CL shows a rotation of $46 \degr \pm 9 \degr$ with respect to the measured position angle of the radio jet and previous marginal results by PolarLight. This may suggest a variation in the polarization angle related to the source state, which is linked to the variation of corona geometry as reported by \ixpe observations of Z sources, or possibly to relativistic precession.}

\FullConference{%
  Frontier Research in Astrophysics, IV
  09-14 September 2024\
 Mondello, Palermo, Italy
}

\begin{document}
\maketitle

\section{Introduction}
On June 18 1962, a team led by Riccardo Giacconi found the first evidence of X-rays from a source outside the solar system \cite{Giacconi62}, using an Aerobee-sounding rocket launched at the White Sands Missile Range (New Mexico). This source, later identified as Scorpius (Sco) X-1, was exceptionally bright and dominated the X-ray sky.  In contrast, no remarkably bright source in visible light or radio was present at its position. This was extraordinary for the knowledge of the time because the only known astrophysical source that emitted X-rays was the Sun, for which the intensity of the X-ray radiation is about one millionth of the intensity of visible light, while for \sco the intensity of X-ray radiation is a thousand times higher than the intensity of visible light \cite{Giacconi03}. Therefore, the mechanism that emitted X-rays was utterly different from that of the Sun: a new astrophysical object was discovered. The emission of X-rays from \sco was confirmed a year later by another proportional counter on board an Aerobee-sounding rocket \cite{Bowyer64a}, followed by the discovery of X-ray emission from the Crab Nebula \cite{Bowyer64b,Bowyer64c}. X-ray astronomy was born, allowing the discovery of several classes of new astrophysical sources completely different from the ones known by observing the sky at other wavelengths. X-ray astronomy allows the study of environments dominated by extreme magnetic and/or gravitational fields, which originate violent phenomena, giving rise to non-thermal high-energy radiation, or to strong asymmetric jets of matter, or to rapidly increasing luminosity of the sources themselves.

With the dawn of X-ray astronomy, it was immediately apparent that X-ray polarimetry was a powerful tool to study the universe. In fact, such extreme processes also leave a distinct signature on the polarization of the X-ray radiation, which can probe the emission mechanism and the geometry of the emitting regions at an angular scale much smaller than the available resolution of X-ray telescopes. Several attempts were made to measure X-ray polarization, but, unfortunately, for more than 40 years, due to technological limitations, the only available significant detection of astronomical X-ray polarization was for the Crab Nebula \cite{Weisskopf76, Weisskopf78}, obtained in 1976 by the Bragg X-ray polarimeter on board the OSO-8 satellite and other marginal detections were reported in hard X-ray \citep{Dean08, Chauvin17} Finally, in 2021, the Imaging X-ray Polarimetry Explorer (IXPE), the first mission fully dedicated to measuring the X-ray polarization in the 2--8\,keV energy band \cite{Soffitta21, Weisskopf22} was launched. IXPE was able to measure the X-ray polarization of several classes of astrophysical sources with high significance. This means that nowadays, thanks to \ixpe, X-ray polarization is a possible investigation tool for potentially any X-ray source.

\begin{figure}[tb]
    \centering
    \includegraphics[width=0.5\linewidth]{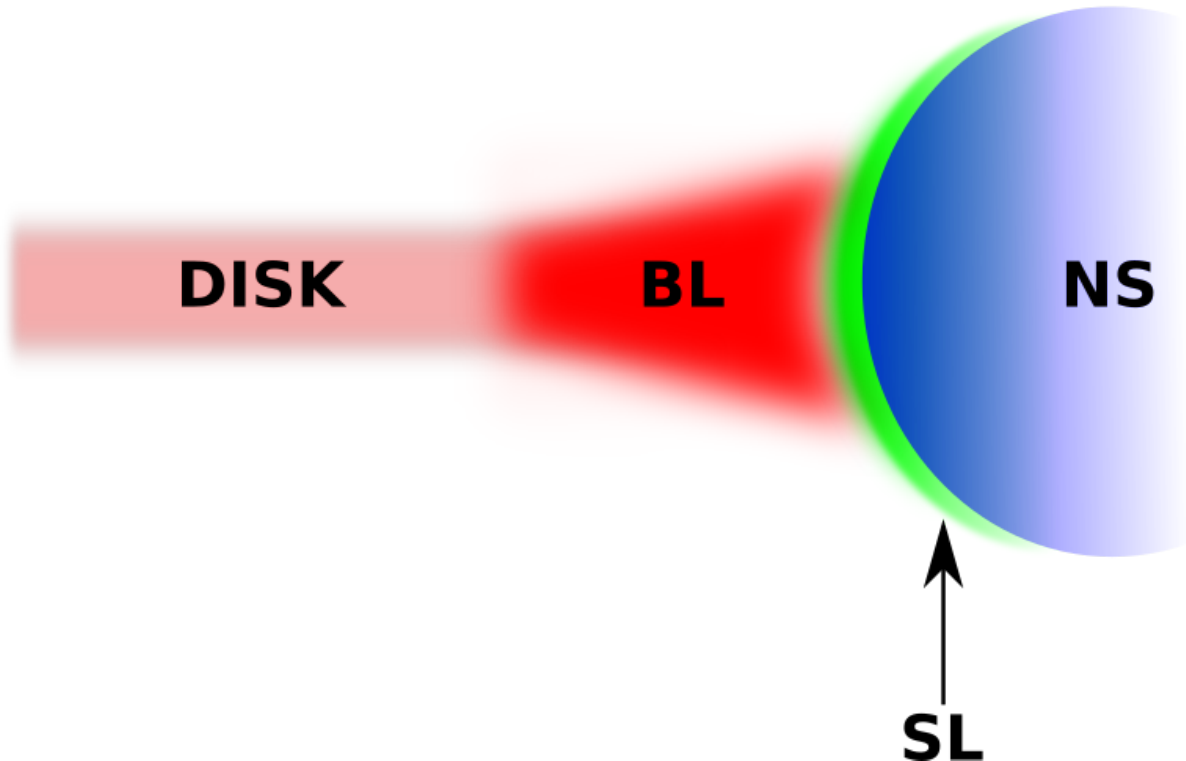}
    \caption{A simplified sketch (not in scale) of the geometry of the inner disk, BL, SL and NS surface}
    \label{fig:sketch_BL_SL}
\end{figure}

Today, we know that Sco X-1 is the brightest persistent extrasolar X-ray source. It is a weakly magnetized neutron star (WMNS) within a low-mass X-ray binary (LMXB) system. Through Roche lobe overflow, WMNS accretes matter from a companion star, with a mass of less than one solar mass, forming an accretion disk around it \cite{Bahramian22}. The NS surface stops the accelerated matter from the inner disk, creating a boundary layer (BL) coplanar with the disk \cite{Shakura88,Popham01}, while the gas forms a spreading layer (SL) as it moves along the NS surface to higher latitudes \cite{Inogamov1999, Suleimanov2006}. A simplified sketch of the BL and SL is reported in Figure~\ref{fig:sketch_BL_SL}. Other possible components that are expected in these systems are an accretion disk extended corona and the presence of a wind above the disk.
WMNSs are classified as Z or atoll sources based on their tracks in the X-ray hard-color/soft-color diagram (CCD) and their timing characteristics \cite{hasinger89, vanderklis89}. Z sources are more luminous than the atoll sources, close to the Eddington luminosity, and show a Z-like track in the CCD with horizontal branch (HB), normal branch (NB), and flaring branch (FB). Furthermore, a hard apex (HA) and a soft (SA) apex may also be distinguished between these branches \cite{Church12,Motta19}. The different branches correspond to different states of the source and different mass accretion rates. Based on their different timing, track patterns, and distinct flare characteristics, Z-sources are further divided into Cyg-like and Sco-like sources \cite{Kuulkers94,Kuulkers97}; the Cyg-like sources have weak flaring activities, a clearly distinguishable HB, and typically a higher inclination or magnetic field compared to Sco-like sources \cite{Kuulkers97, Psaltis95}. Up to now, we know three persistent Cyg-like sources (\mbox{GX~340+0}, \mbox{Cyg~X-2}, and \mbox{GX~5$-$1}) and three Sco-like sources (\mbox{Sco~X-1}, \mbox{GX~349$+$2}, and \mbox{GX~17$+$2}). 

Since they are very bright X-ray sources, they have been studied in great detail since the beginning of X-ray astronomy. This was done using spectroscopy and timing to try to figure out the emission mechanism and the possible geometry of the inner disk, BL/SL and NS surface. Typically, a hard Comptonized component from the BL/SL region and a softer thermal component from the accretion disk and/or NS surface may be used to characterize the spectrum of these sources. The spectrum often also shows an iron emission line at ${\sim}6.4$\,keV, which indicates reflection from the cold matter of the inner accretion disk. The study of the reflection features is a very interesting tool for determining the inner radius of the accretion disk and to shed light on the area close to the compact object. For the modeling and the implications of the reflection in NS-LMXBs, see, e.g., Ref.~\cite{Ludlam24} and reference therein. WMNSs do not show any pulsations but can show quasi-periodic oscillations (QPOs). The timing properties of the QPOs, combined with the spectral studies, revealed that the strongest variability is associated with a harder spectral component correlated with the BL/SL \citep{gilfanov2003,Revnivtsev06, Revnivtsev13}. Unfortunately, even if the timing properties are combined with the spectral analysis, the geometry of the various emitting regions is not entirely revealed due to the degenerate spectral features of these regions.

As reported before, X-ray polarization is an effective method for determining the geometry and emission processes of the accretion flow region. In fact, the BL/SL, the extended corona, and the accretion disk show distinctive X-ray polarization characteristics \cite{Loktev22, Poutanen23, Bobrikova24SL, Farinelli24, Tomaru24}, which can now be measured thanks to \ixpe. In the first 3 years, \ixpe observed 15 WMNSs, in particular five atoll sources: \mbox{GS~1826$-$238} \citep{Capitanio23}, \mbox{GX~9+9} \citep{Ursini23}, \mbox{4U~1820$-$303} \citep{DiMarco23b}, \mbox{4U~1624$-$49} \citep{Saade24}, \mbox{Ser X-1} \citep{Ursini24}, and \mbox{GX~3$+$1} \citep{Gnarini24}; four Z-sources: \mbox{Cyg X-2} \citep{Farinelli23}, \mbox{GX~5+1} \citep{Fabiani24}, \mbox{Sco X-1} \citep{LaMonaca24}, and \mbox{GX~340+0} \citep{LaMonaca24b}; two persistent peculiar sources: \mbox{Cir X-1} \citep{Rankin24} and \mbox{GX~13$+$1} \citep{Bobrikova24a,Bobrikova24b}; and the Z-atoll transient \mbox{XTE J1701$-$462} \citep{Cocchi23}. The average polarization measured for the atoll sources is 1--2\%, in some cases a dependence with energy \cite{Ursini23, DiMarco23b, Saade24} is also observed. In particular, \mbox{4U~1820$-$303} shows an exceptional  $\sim$10\% PD above 7\,keV \citep{DiMarco23b}. For Z-sources, IXPE measured a higher average polarization that varies with the branch and the energy: 4--5\% if the source is in the HB (\mbox{GX~5$-$1} \citep{Fabiani24}, \mbox{XTE J1701$-$462} \citep{Cocchi23} and  \mbox{GX 340+0} \citep{LaMonaca24b}); below 2\% if the source is in the NB or FB (\mbox{Cyg X-2} \citep{Farinelli23} and \mbox{Sco X-1} \citep{LaMonaca24}). Regarding the polarization angle (PA), IXPE measured a PA aligned with the radio jet's direction in \mbox{Cyg X-2}, while PA variations with time/state were observed in \mbox{Cir X-1} \citep{Rankin24}, \mbox{GX~13$+$1} \citep{Bobrikova24a,Bobrikova24b} and \mbox{XTE J1701$-$462} \citep{DiMarco24, Zhao24}.

Due to their brightness, WMNSs are the best candidates not only for X-ray astronomy but also for X-ray polarimetric measurement, and therefore \sco, being the brightest X-ray source, was one of the first to also be observed by the OSO-8 satellite in 1977 \cite{Long79}. Unfortunately, this measure results in only an upper limit at 2.6\,keV, while at 5.2\,keV they obtained a $\mbox{PD=}1.3\%\pm0.4\%$ and a $\mbox{PA}=57\degr \pm 6\degr$, slightly above $3\sigma$ of confidence level (CL) with an exposure time of ${\sim}$15 days. In 2022, another attempt \cite{Long22} was performed by PolarLight \cite{Feng19},  a Chinese CubeSat mission equipped with a photoelectric X-ray polarimeter, similar to the ones on board of IXPE. This instrument is capable of detecting polarization in the 2--8\,keV energy band without any focusing system, and it is characterized by a very small effective area, limiting the observations to very bright sources with very long exposure time. It observed \sco for ${\sim}$322\,days indicating variation of PD with energy and reporting in the "high-flux" state and in 4--8\,keV a $\mbox{PD=}4.3\%\pm0.8\%$ at a $\mbox{PA=}53\degr \pm 5\degr$ at 5$\sigma$\,CL \citep{Long22}. The most notable aspect of this result is that the PA agrees with the previous observation by OSO-8 and is aligned with the position angle measured for the radio jet. This alignment was explained by \cite{Long22} as an optically thin vertically extended corona.
\sco is one of the very few NS for which radio observations, using very-long-baseline interferometry, were able to spatially resolve the jet at submilliarcsecond scales, measuring a position angle of ${\sim}54\degr$ (measured from north to east) \cite{Fomalont01, Fomalont01b}. It is very interesting to have a measure of the direction of the radio jet because it is almost perpendicular to the plane of the orbit and, therefore, to the disk. Consequently, if the PA is aligned with that direction, it is also perpendicular to the disk. This PA direction is in agreement with a corona geometry like a vertically extended optically thick corona (i.e., the SL) that lies between the inner disk and the NS surface; on the contrary, if there is an optically thick accretion disk corona above the disk (or a BL), the PA should be aligned with the disk direction and perpendicularly to the radio jet's direction. Unfortunately, both observations prior to \ixpe are difficult to use for a direct comparison with the new one because those observatories were not capable to
correlating the X-ray polarization with the state of the source and, for some WMNSs, \ixpe showed variations of the polarization with the state of the source \cite{Cocchi23, Fabiani24, Rankin24, Bobrikova24a}.

\section{IXPE observation of Sco X-1: polarimetric analysis }
\subsection{X-ray simultaneous observational campaign of \sco}
Sco X-1 was observed by \ixpe on 2023 August 28 for approximately 24 ks per each IXPE telescope. Strictly simultaneous observations were conducted by \nicer \citep{Gendreau16}, \nustar \citep{Harrison13}, and \hxmt \citep{Zhang14, Zhang20}. A detailed description of this X-ray observation campaign is reported in Ref.~\cite{LaMonaca24}. The goal of this simultaneous campaign was to correctly identify the state of \sco and to have an improved broad-band spectral model compared to the one that could be obtained using only \ixpe due to the better spectral capabilities of these observatories. Moreover, the observation at higher energy, above 10\,keV, is fundamental for better constraining the spectral reflection features.

\begin{figure}
    \centering
    \includegraphics[width=0.9\linewidth]{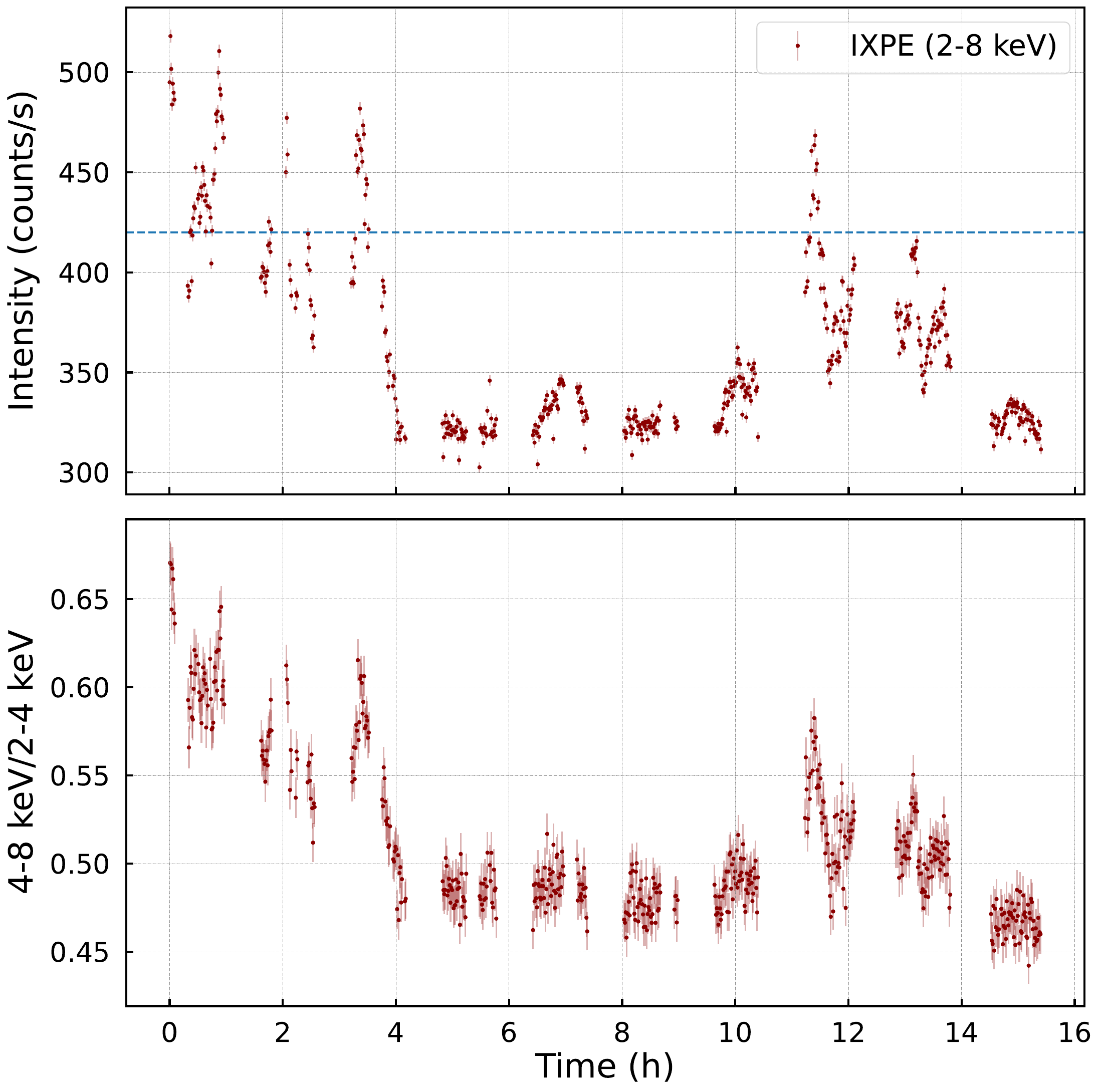}
    \caption{\textit{Top panel}: \ixpe light curve with the corresponding HR binned in 100 s intervals. The blue dotted line identifies the threshold for the high and low flux states. \textit{Bottom panel}: \ixpe HR is obtained as the ratio of the IXPE counting rates in the 4–8 and 2–4 keV energy bands.}
    \label{fig:ixpe_lc_hr}
\end{figure}

\begin{figure}
    \centering
    \includegraphics[width=0.7\linewidth]{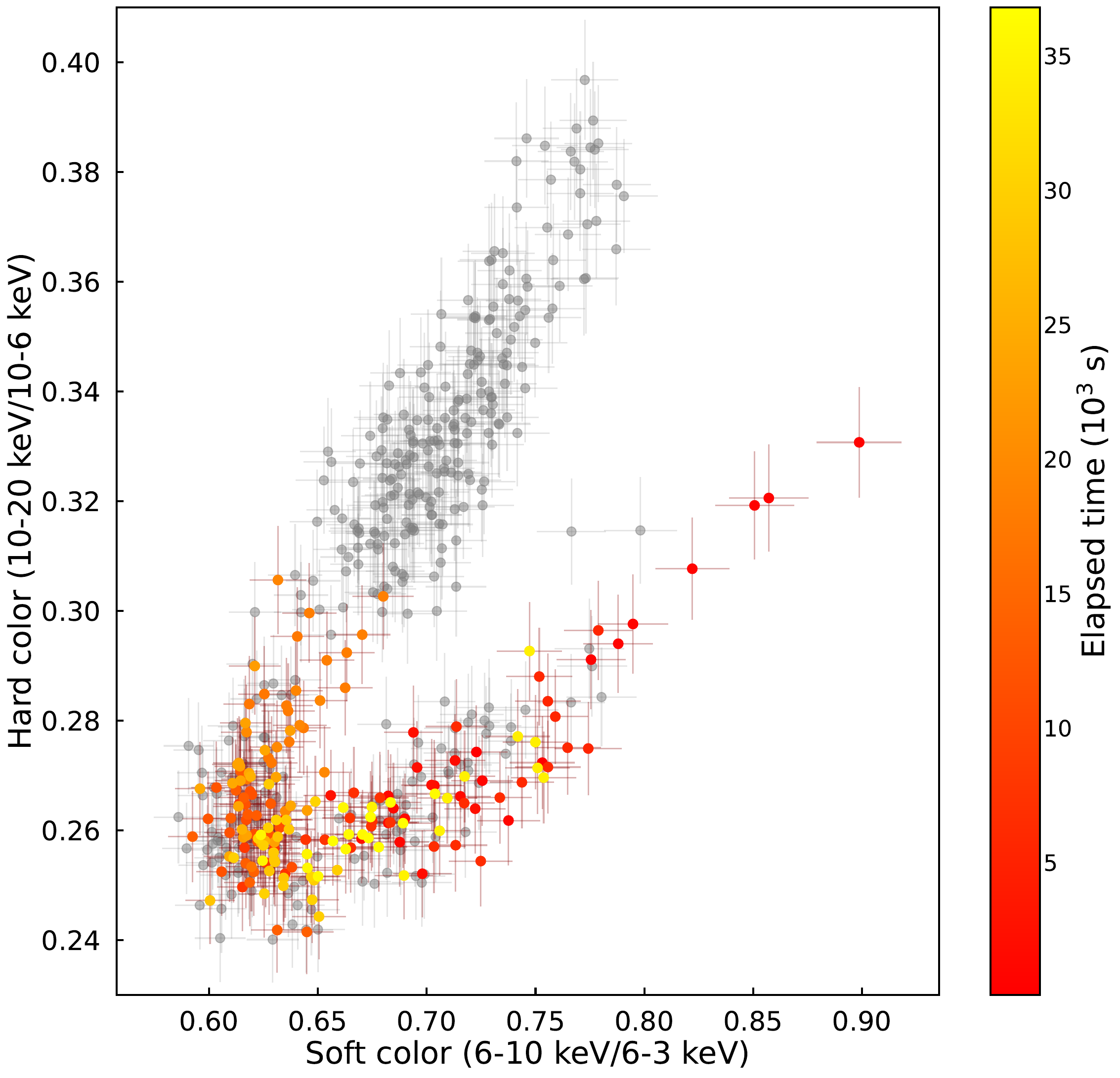}
    \caption{CCD of \sco from \nustar data archive (gray points). Overlapping colored points identify the simultaneous \nustar observation, with the color changing from red to yellow with the time elapsed since the observation began.}
    \label{fig:NuSTAR_CCD}
\end{figure}

Due to the \sco high brightness, especially in the soft band, \ixpe observation of \sco was performed using the gray filter, a partially opaque absorber, which strongly reduces the incident flux, especially below 3\,keV. This filter is a 76.2\,$\mu$m foil Kapton with 100\,nm of Al, and it is included in the \ixpe detector's filter calibration wheel \citep{Ferrazzoli20, Soffitta21, Weisskopf22} to allow the observation of very bright sources, reducing the count rate to a level acceptable for the \ixpe detector’s dead-time. In Figure~\ref{fig:ixpe_lc_hr}, the \ixpe light curve of \sco is shown (\textit{top panel}) with the time evolution of the harness ratio (\textit{bottom panel}). Some flaring activity is evident, as the rapidly increasing count rates in the light curve show in some intervals.
The light curves, the hardness ratio, and the CCD, presented here, were obtained using \textsc{stingray}, which is developed in \textsc{Python} for the purpose of performing spectral timing analysis \citep{stingray2,stingray1,Bachetti24}.
This flaring activity is reported in all the X-ray telescopes observing \sco jointly with \ixpe (see Figure~1 in Ref~\cite{LaMonaca24}). The \ixpe hardness ratio does not allow for a clear identification of the state of \sco; therefore, we computated a CCD with the archived \nustar observations, and we compared it with the new \nustar observation simultaneous with the \ixpe one (see Figure~\ref{fig:NuSTAR_CCD}).
From the \nustar CDD, we were able to draw the conclusion that \sco was going from the FB to the NB and again through the FB at the end of the observation, in perfect agreement with the \ixpe light curve. Therefore, the \ixpe observation was dominated by the SA with short periods in the FB. This is also confirmed by the timing analysis of \nicer simultaneous observation \citep{LaMonaca24}, which detects a QPO at ${\sim}$6\,Hz, which is the frequency expected in the SA of \sco  \citep{Casella06}.

\subsection{Polarimetric analysis}

To perform a model-independent analysis of the X-ray polarization without making any assumptions on the spectral model, it is used the \texttt{pcube} algorithm based on the approach of Ref.~\cite{Kislat15}, inside \textsc{ixpeobssim} \citep{Baldini22}, a dedicated software to analyze and simulate \ixpe observations. This analysis method with the gray filter does not accurately account for the telescope response matrices at low energy \citep{LaMonaca24, Veledina23}. This causes an overestimation of polarization below~3\,keV, compared to the results obtained by spectropolarimetric analysis with XSPEC \citep{Arnaud96} and so the model-independent analysis is precautionary performed only above 3\,keV. A PD of $1.08\%\pm0.15\%$ at a PA$=10\degr\pm4\degr$ at 68\%\,CL were measured on the entire observation in the 3--8\,keV with a detection significance of $\sim$7$\sigma$ CL. The polarimetric analysis is performed without background rejection or subtraction due to the source's brightness, following the prescription in Ref.~\cite{DiMarco23a}.

\begin{figure}
\centering
	\includegraphics[width=0.8\linewidth]{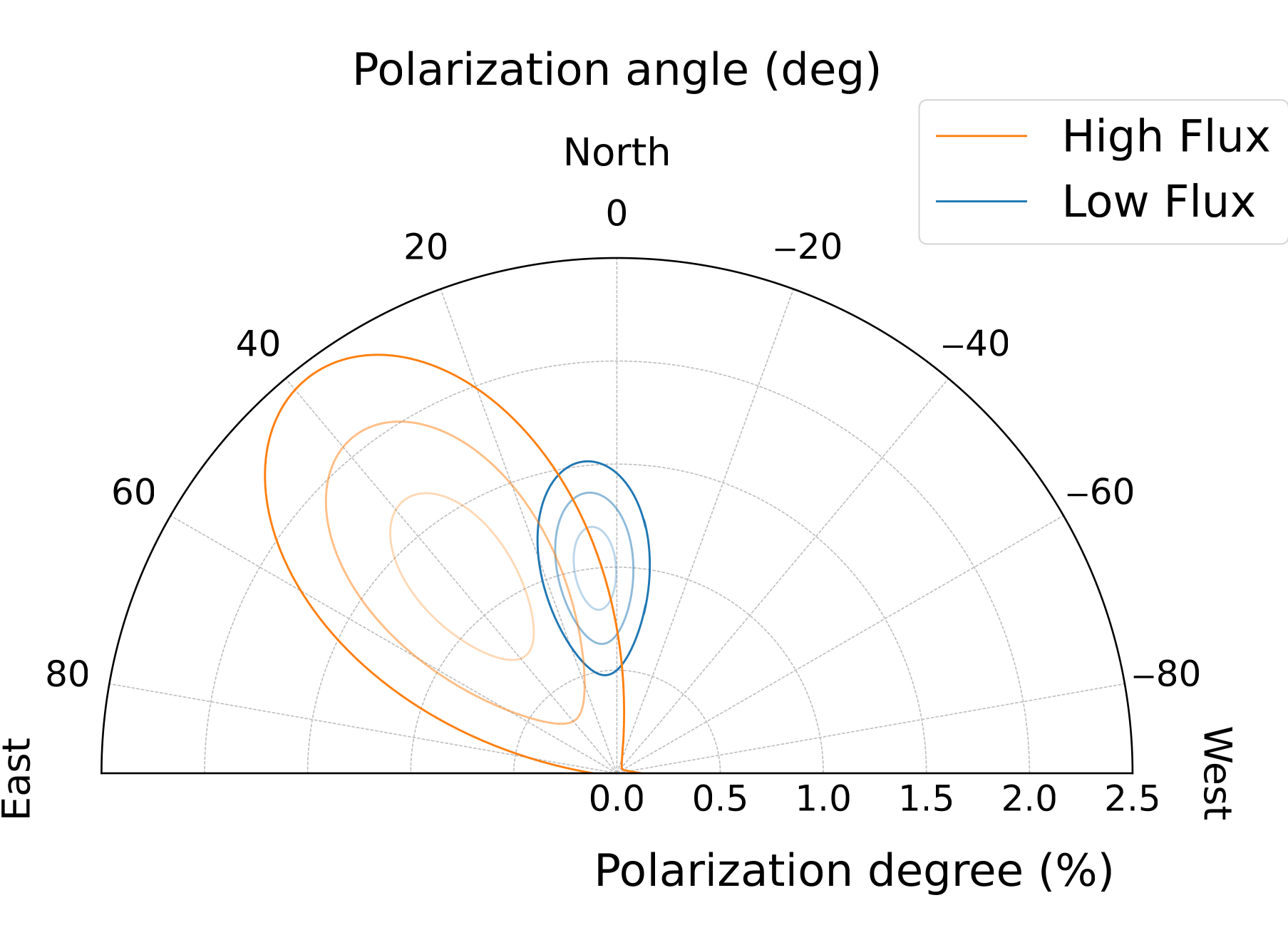}\\
\caption{Polar plot of the PD and PA for \sco in the 3--8\,keV energy band obtained by \textsc{ixpeobssim} when the observation is divided in high and low flux state.  The contours are at 50\%, 90\%, and 99\%~CL from the innermost to the outermost. \label{fig:contour_flare_430cts}}
\end{figure}

Considering the presence of flaring activity during the \ixpe observation, an attempt to investigate the polarization properties of flaring and non-flaring states was performed (see Figure~5 in Ref.~\cite{LaMonaca24}) on the basis of the \nustar time intervals, finding that the polarization is compatible within 90\%\,CL. A further attempt was performed by using only the \ixpe observation and dividing it into high and low flux state. Here, it is reported the result for a threshold of 430~counts/sec (see the blue dotted line in Figure~\ref{fig:ixpe_lc_hr}). The resulting contour plots are reported in Figure~\ref{fig:contour_flare_430cts}. The two states show a polarization which is compatible within 90\% CL (similar results were obtained with different threshold values), while at 99\%\,CL the high flux state polarization is not significant. Varying the threshold value, the high flux contour plots are larger and less significant (higher threshold) or more significant but much more compatible with the low flux state. Thus, confirming the result reported in \cite{LaMonaca24}, the present data do not allow for a claim for a significant polarization variation with the flux.

\begin{figure}
    \centering
    \includegraphics[width=0.9\linewidth]{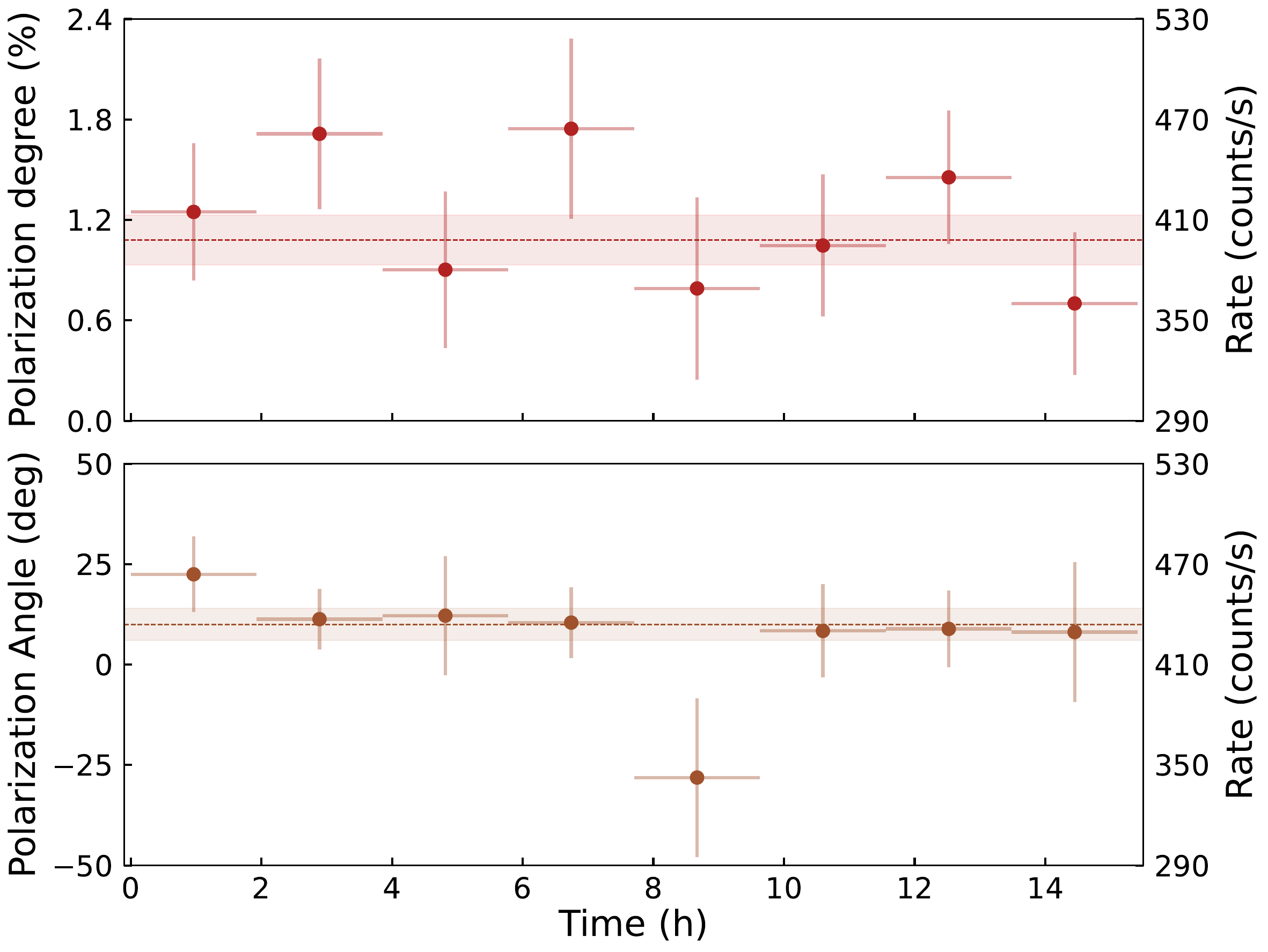}
    \caption{PD (top panel) and PA (bottom panel) resolved in time during the observation of \sco, overlapped to the \ixpe light curve (gray points). The observation is divided into 8 equal time bins of ${\sim}1.9$\,hours. Errors are reported at 68\%\,CL. The colored dashed lines indicate the average-time PD and PA values in the whole observation in the 3--8\,keV energy band, and the colored pink areas are the 68\%\,CL on those values.}
    \label{fig:PD_PA_vs_time}
\end{figure}

Since some WMNSs showed variation with time (see, e.g., \mbox{GX~13$+$1} \citep{Bobrikova24a,Bobrikova24b} and \mbox{XTE J1701$-$462} \citep{DiMarco24, Zhao24}), the \sco observation is divided in into 8 equal time bins of ${\sim}1.9$\,hours and the PD and PA are obtained for each interval in the 3--8\,keV energy band. The results are shown in Figure~\ref{fig:PD_PA_vs_time}. No significant variations are observed for PD and PA with time and/or with the flux of the source because all PD values are compatible within 68\%\,CL, and the same for the PA values with the exception of the fifth time bin that is compatible within $2\sigma$.   

\subsection{Spectropolarimetric analysis}\label{sec:spec_pol}

To model the spectrum of \sco during the IXPE observation, data from NICER in the 1.5--12 keV energy band, NuSTAR in the 3--40\,keV energy band, Insight-HXMT in the medium (8--30\,keV) and high (30--40\,keV) energy bands, and IXPE in the 2--8\,keV energy band were used. An extended description of the spectral and spectropolarimetric analysis is reported in Ref.~\citep{LaMonaca24}. The spectral model exhibits a continuum described by \texttt{diskbb} \citep{Mitsuda84, Makishima86}, a soft multicolor disk emission, and by a  \texttt{nthcomp} \citep{Zdziarski96, Zychi99} for the hard Comptonized component. Additionally, a broad Gaussian is included in order to account for the reflection feature due to the presence of an iron fluorescence line. In the \ixpe data, the reflection features are hard to detect due to the worse spectral capabilities of this satellite compared to the other observatories. Thus, this simplified model for the reflection is sufficiently appropriate for carrying out the spectropolarimetric analysis. The simultaneous observations were fitted with the \textsc{xspec} model \texttt{tbabs*(diskbb+nthcomp+gauss)}, named Model~A, where the interstellar medium absorption was fixed at the galactic value of $0.15 \times 10^{22}$ cm$^{-2}$ \citep{Ding23, Kalberla05, HI4PI}. Using the spectral parameters fixed at the best value for this model (see Model A of Table 3 in Ref.~\citep{LaMonaca24}), the $I$, $Q$, and $U$ spectra were simultaneously fitted using the \textsc{xspec} model \texttt{tbabs*(polconst*(diskbb+nthcomp)+gauss)}, where a \texttt{polconst} model (which describes a constant polarization with the energy) is associated to the continuum, while a \texttt{polconst} with zero amplitude is associated to the Gaussian line, which is expected to be almost unpolarized, due to an isotropic fluorescence process \citep[see, e.g.,][]{Churazov02, Ingram23}. Therefore, in the whole 2--8\,keV energy band, the
PD was found to be 1.0\% $\pm$ 0.2\%  at a PA of $8\degr \pm 6\degr$ with a $\chi^2$/dof = 1371/1341=1.02. The polarimetric results are reported in Figure~\ref{fig:ixpe_contour_radio_jet}~(\textit{top panel}) as a polar plot in the PD/PA plane. The \textsc{xspec} spectropolarimetric results confirm the model-independent analysis, providing evidence that the study carried out using the \texttt{pcube} algorithm above 3\,keV was accurate and allowing the extension of the polarimetric analysis below 3\,keV. A study of the energy-dependent polarization was performed in two different ways. First, the analysis was carried out by splitting the IXPE energy band into energy bins of 1 keV and performing the fit with the previous spectropolarimetric model in each energy bin. 
\begin{figure}
    \centering
    \includegraphics[width=0.8\linewidth]{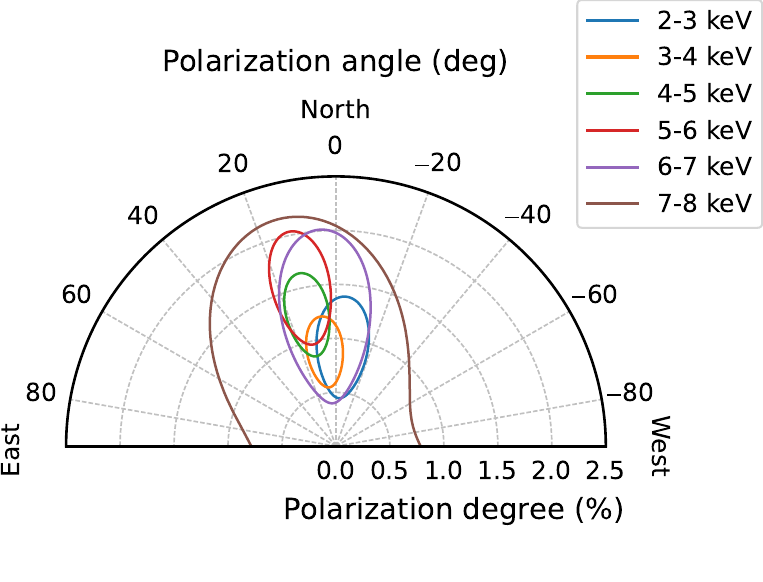}
    \caption{Polar plot of the PD and PA for \sco, using the \textsc{xspec} model \texttt{tbabs*(polconst*(diskbb+nthcomp)+gauss)} in each 1\,keV energy bin in the \ixpe nominal energy band; contours are reported at 68\%\,CL. No energy trend for the polarization is present. See also Table~2 and Figure~4\,(c) in Ref.~\citep{LaMonaca24} to compare these results with the model-independent analysis in 3--8\,keV energy band.}
    \label{fig:contours_xspec_1keV}
\end{figure}
\begin{figure}[h!]
\centering
	\includegraphics[width=0.8\linewidth]{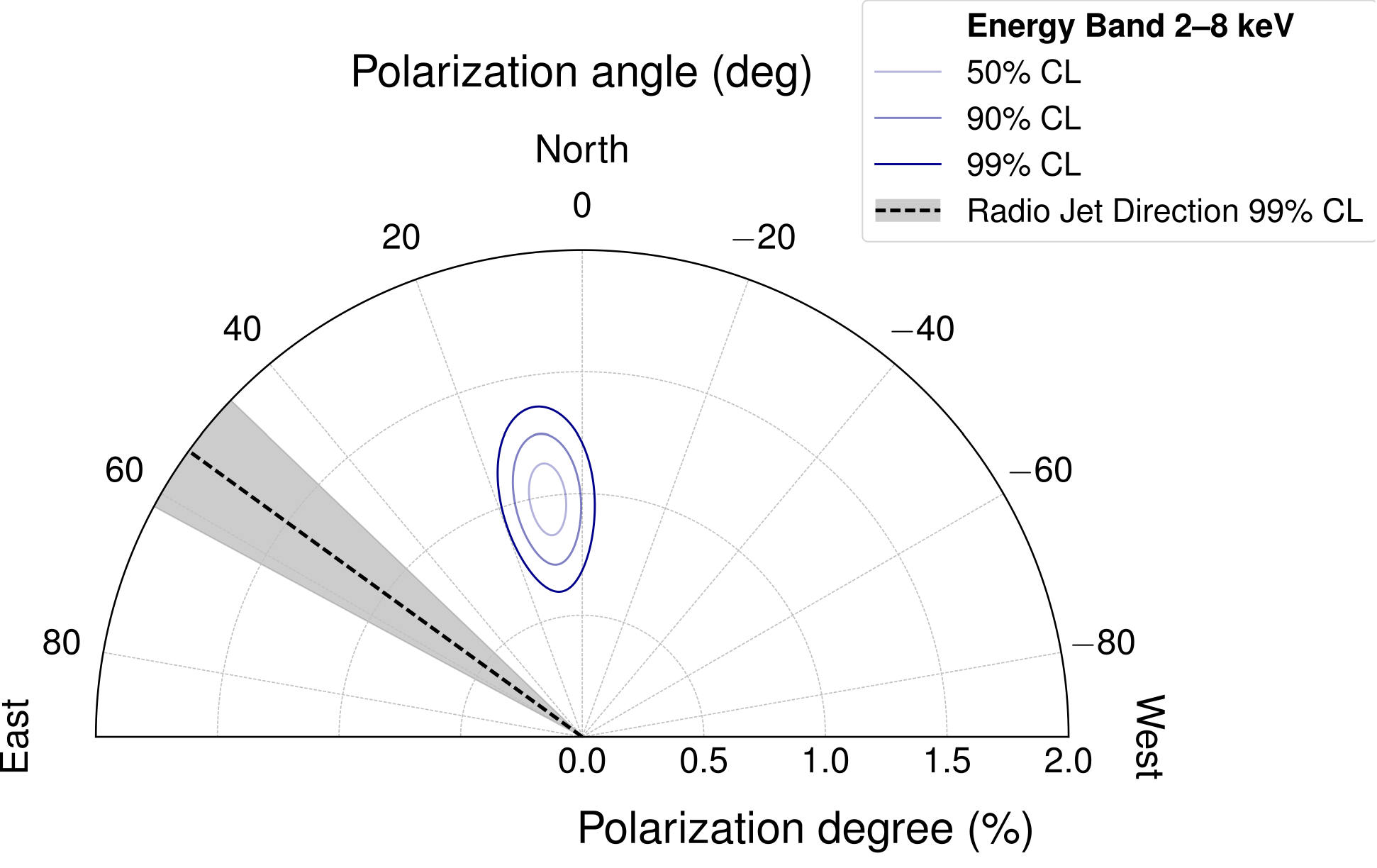}\\
    \vspace{0.5cm}
   \includegraphics[width=0.8\linewidth]{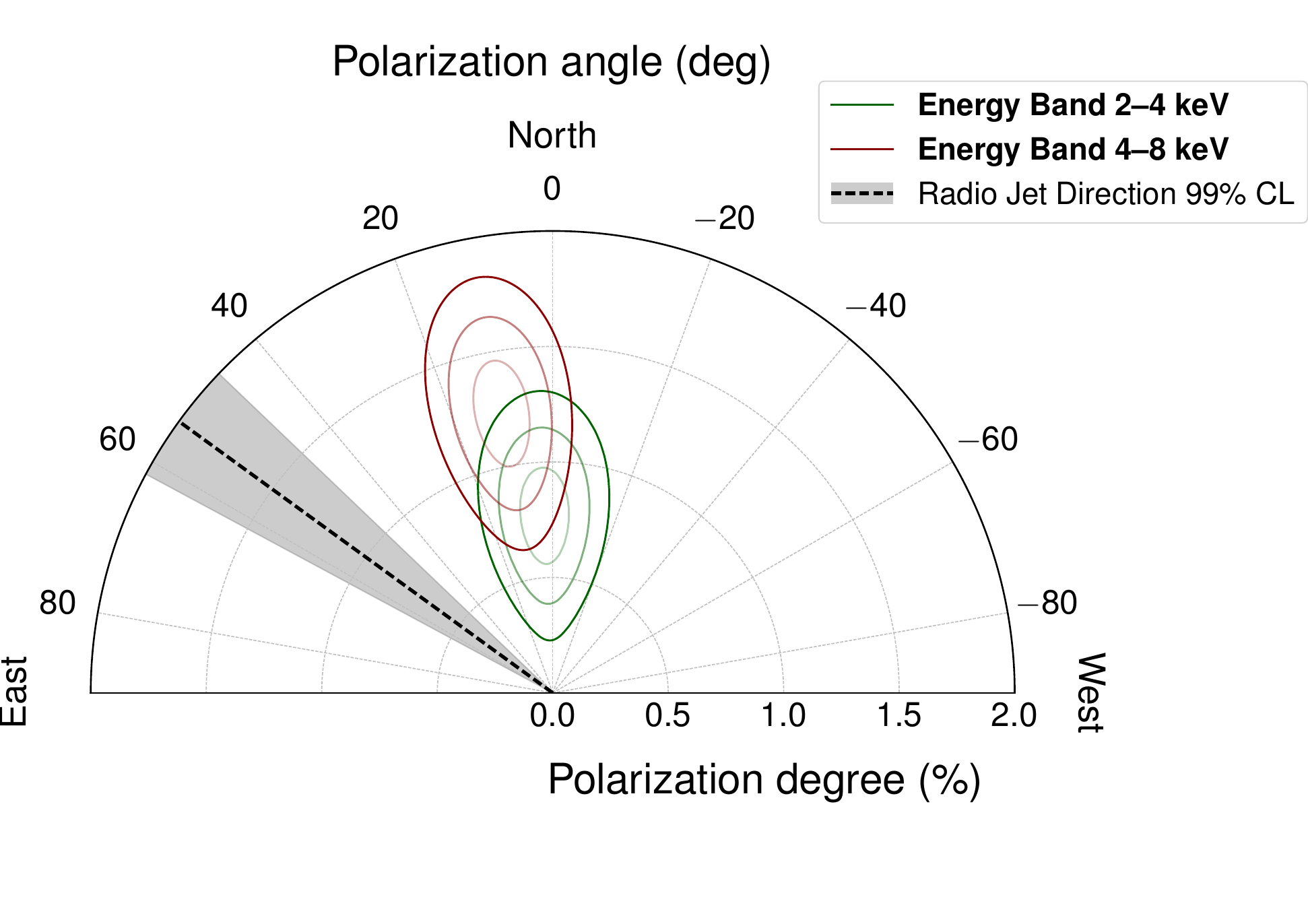}
\caption{Polar plot of the PD and PA for \sco obtained by \textsc{xspec} in the nominal \ixpe energy band (\textit{top panel}) and in the  2--4\,keV and 4--8\,keV energy bands (\textit{bottom} panel) \citep{LaMonaca24c}. The contours are at 50\%, 90\%, and 99\%~CL from the innermost to the outermost. The direction of the radio jet is reported at 99\%~CL as measured by Refs.~\cite{Fomalont01,Fomalont01b}.  \label{fig:ixpe_contour_radio_jet}}
\end{figure}
The contour plots for each energy bin are reported in Figure~\ref{fig:contours_xspec_1keV}, and they are all compatible within 68\%\,CL (see also Table~2 in Ref.~\cite{LaMonaca24}). Spectropolarimetric analysis using the \texttt{pollin} model in \textsc{xspec} (which describes a polarization that can vary linearly with the energy) is performed, finding that the PD and PA slopes are compatible with zero \citep{LaMonaca24}. Thus, in the \ixpe nominal energy range, there are no significant variations for PD or PA, in agreement with the results obtained by performing the spectropolarimetric analysis in each energy bin of 1\,keV.

Using the model \texttt{tbabs*(polconst*diskbb+polconst*nthcomp+gauss)}, with the two constant polarizations free to vary independently without any tieing, the spectropolarimetric analysis also makes it possible to investigate distinctly the polarization of the soft and the Comptonized components. These results are reported in Table~\ref{tab:specpol} as model~A. Unfortunately, it was impossible to determine the PD of the disk component, setting only an upper limit.

The simultaneous X-ray spectra can also be fitted with a model that fully describes all the reflection features as \texttt{relxillNS} \citep{Garcia22}, which can be added to the continuum using the model \texttt{tbabs*(diskbb+comptt+relxillNS)}. In this scenario, a thermal blackbody, originating from the NS or SL/BL surface, illuminates the cooler accretion disk and produces the continuum of the reflection spectrum. It was not possible to constrain the inclination in the reflection model with the present data; therefore, the value is fixed at 44\degr,~as determined by radio observations \cite{Fomalont01,Fomalont01b}.
The best-fit parameters of the spectral fit are reported in Table~2 in Ref.~\citep{LaMonaca24}, where this model is referred to as Model~C, and the reduced $\chi^2$ is improved compared to the simple model with the Gaussian as well as when compared to a fit in which the \texttt{diskline} model was used. In all the different models, the component that dominates the flux is the Comptonized one, and the contribution of the flux of the reflection to the overall one is around 10\%.
The reflection fitting allows for setting an upper limit on the inner disk truncation radius at $9\,R_{\rm g}$ ($1.50\,R_{\rm ISCO}$), in agreement with the one derived using the \texttt{diskline} model \citep{LaMonaca24}, and showing the inner radius of the disk is close to the $R_{ISCO}$. By associating a polarization constant to each spectral component of the model with the reflection, it is not possible to constrain the polarization of the reflection component, neither of the soft and hard components, but only to set upper limits. The results of this spectropolarimetric analysis are reported in Table~\ref{tab:specpol} as model~C1. Therefore, a clear disentangling of these three spectral components is not possible. To constrain the polarization of the reflection, the polarization of the disk component is fixed at 1.1\% as predicted by Refs. \cite{Chandrasekhar1960, Sobolev63, Loktev22} and the Comptonized one is assumed unpolarized \citep{st85}, considering that the inclination of \sco is ${\sim}44\degr$. These results are reported as Model~C2 in Table~\ref{tab:specpol}. If we assume a higher polarization for the Comptonized component, the polarization of the reflection component decreases, showing a degeneracy of the PD of these two components, as reported also in Ref.~\cite{LaMonaca24b}. The assumption of an unpolarized Comptonization allows for obtaining the maximum PD for the reflection component. 
\begin{center}
\begin{table}[h]
\centering
\caption{Best-fit parameters for the
spectropolarimetric analysis with the different models reported in Refs.~\citep{LaMonaca24,LaMonaca24c}. Errors are at 90\%~CL\label{tab:specpol}}%
\tabcolsep=10pt%
 \begin{threeparttable}
\begin{tabular*}{30pc}{@{\extracolsep\fill}llccc@{\extracolsep\fill}}
\toprule
\textbf{Component} & & \textbf{Model A} & \textbf{Model C1} & \textbf{Model C2} \\
\midrule
\texttt{diskbb} & PD (\%) & $<3.2$ & $<1.9$ & $1.1^{\dagger}$\\
    & PA ($\deg$) & --- & ---  & $-40^{\dagger}$ \\ \hline
\texttt{nthcomp} & PD (\%) & $1.3 \pm 0.4$ & $<8.2$ & $0^{\dagger}$\\ 
    & PA ($\deg$)  & $14\pm8$ & --- & -- \\ \hline
\texttt{rellxillNS} & PD (\%) & -- & $<66\%$ & $14\pm5$ \\ 
    & PA ($\deg$) & --  & --- & $15\pm7$\\
\bottomrule
\end{tabular*}
\begin{tablenotes}
\item[$^{\dagger}$] Fixed
\end{tablenotes}
   \end{threeparttable}
\end{table}
\end{center}

\section{Discussion and Conclusion}

\ixpe observed \sco in a strictly simultaneous X-ray observation campaign with also included \nicer, \nustar, and \hxmt, which allowed broad-band spectral modeling and identification of the state of the source during the observation. \ixpe observed \sco mainly in the SA state with short flaring periods. The polarization of \sco was measured with a high significance, about $7\sigma$~CL. The polarization of the SA state and the flaring period showed no significant variation at 90\%\,CL.

Assuming the spectral model A in Section~\ref{sec:spec_pol} and an unpolarized Gaussian line, the spectropolarimetric analysis allowed to measure a PD of $1.0\% \pm 0.2\%$ at a PA of $8\degr \pm 6\degr$ at 90\%~CL in the 2--8\,keV energy band. An energy-resolved spectropolarimetric study with 1\,keV energy binning for the polarization (see Figure~\ref{fig:contours_xspec_1keV}) or using the \textsc{xspec} \texttt{pollin} model showed no significant variation, confirming the result obtained by the model-independent analysis in 3--8\,keV \citep{LaMonaca24} and in contrast to other \ixpe results for atoll sources (see, e. g., \mbox{4U~1820$-$303} \citep{DiMarco23b}) and Z sources (see, e. g., \mbox{GX~340+0} \citep{LaMonaca24b},  \mbox{GX~5+1} \citep{Fabiani24} and \mbox{XTE J1701$-$462} \citep{Cocchi23}).  The constant polarization with energy can imply a sandwich-like corona geometry (see the geometry for case~C in Ref.~\citep{Poutanen23}). The spectropolarimetric analysis allowed for determining the polarization of the soft disk component and the hard Comptonized one. For the disk, the PD results in an upper limit below 3.2\% at 90\%~CL, in agreement with the theoretical predictions ${\sim} 1.1\%$ for an electron-scattering dominated and optically thick accretion disk \citep{Chandrasekhar1960, Sobolev63, Loktev22}, and for the Comptonized component a PD of ${\sim}1.3$\%, in agreement for the PD of slab of Thomson optical depth of $\sim$7 and an electron temperature of $\sim$3\,keV \citep{st85}, where the optical depth and the electron temperature are obtained by the spectral fit \citep{LaMonaca24}.

The strictly simultaneous observation of \nicer, \nustar, \hxmt, and \ixpe allowed for a study of \sco's broad-band spectrum. In particular, it is possible to study the reflection component using the \texttt{relxillNS} model in \textsc{xspec}, allowing the inner radius to be constrained with the value of  ${\sim}$18\,km for a NS of 1.4M$_\odot$.  This suggests that the inner radius is close to the NS surface and can be considered as an upper limit on the NS radius. This value is higher than the one obtained for another Sco-like source as \mbox{GX~349$+$2} \citep{Coughenour18}, and in agreement with the results reported for other Cyg-like sources (see, e.g, \mbox{Cyg~X-2} \citep{Ludlam22} and \mbox{GX 340$+$0} in the HB \citep{LaMonaca24b, Li25}).

\textsc{xspec} spectropolarimetric analysis was applied to the reflection model and to obtain a constrained PD for the reflection component of $14\%\pm5\%$ at a PA=$15\degr\pm7\degr$ with error at 90\%~CL (see Model~C2 in Table~\ref{tab:specpol}), fixing the PD of the disk component at  1.1\% \citep{Chandrasekhar1960, Sobolev63, Loktev22} and an unpolarized Comptonized component \citep{st85} as expected considering the inclination of \sco and the values obtained from the fit for $\tau$ and $kT_{e}$ (see Table~2 in \cite{LaMonaca24}). This value aligns with the expected polarization values for cold matter Compton-reflected spectra \citep{Matt93, Poutanen96}.

This highly significant X-ray polarization measurement of \sco by \ixpe enhances the previous attempts by OSO-8 \citep{Long79} and PolarLight \citep{Long22}, which lacked spectrum analysis capabilities and, as reported before, were performed with very long exposure time. These aspects have to be taken into account when those previous observations are compared with that of \ixpe. The upper limit of PolarLight data in the low-flux state aligns with \ixpe results, but when the data are selected in the high-flux and in the 4--8\,keV energy band, it shows a discrepancy in the PA with respect to \ixpe observation. In fact, PolarLight, for this particular selection of the data, detected a PD of ${\sim}4\%$ and a PA of ${\sim}53\degr$ with a ${\sim}5\sigma$\,CL \citep{Long22} and the PA agrees with the one reported by OSO-8 with a lower significance. Moreover, the PA is aligned with the radio jet direction \citep{Fomalont01,Fomalont01b} and, since the radio jet is emitted orthogonal to the orbital plane and the disk, also the PA is orthogonal to the disk itself.
Therefore, the PA observed by PolarLight is compatible with an optically thin vertically extended corona between the disk and NS surface and in contrast with an optically thin accretion disk corona above the disk. In the case we assume an optically thick Comptonizing medium, such as SL/BL, the interpretation is the opposite with polarization vector rotated by 90\degr.  On the contrary, the \ixpe observation indicates a significant misalignment between the measured PA and the radio jet direction reported in Refs.~\citep{Fomalont01, Fomalont01b}, with a rotation of a $46\degr \pm 9\degr$. Figure~\ref{fig:ixpe_contour_radio_jet} shows this misalignment in the 2--8\,keV (\textit{top panel}), which is also present if we select different energy bands (\textit{bottom panel}): 2--4\,keV and 4--8\,keV (the energy band used in the PolarLight analysis). For other sources, \ixpe reported a PA aligned with the radio jet direction for \mbox{Cyg~X-2} \citep{Farinelli23} and for the stellar-mass BHs \mbox{Cyg~X-1} \citep{Krawczynski22}, \mbox{Swfit~J1727.8$-$1613} \citep{Veledina23} and \mbox{GX~339$-$4} \citep{Mastroserio24}, even if radio measurements were not simultaneous. Recently, this variation was also observed for the first time in the NS \mbox{Cir~X-1} \citep{Cowie_eas24,Cowie_RAS24}.  This point has to be considered since some sources exhibit fast variation of the radio jet's direction over a timescale of hours, such as V404 Cygni \cite{Miller-Jones19}, where the direction varies of 36\degr. This variation was linked to the relativistic Lense-Thirring precession of the accretion disk \citep{Stella98}, which may produce a rotation matching the observed value by IXPE. Moreover, \ixpe observations of other WMNSs showed a variation of the PA with the state/time of several tens of degrees (up to ${\sim}70$) \citep{Bobrikova24a, Rankin24, Cocchi23, DiMarco23b}, that are correlated with change of the corona geometry across the different states, resulting in a variation of the PA. This may explain the non-alignment with the previous radio jet's direction measurement.

Until today, we have measured the X-ray polarization of \sco only in the SA state, and a systematic study for possible PA variation throughout the Z branches is not possible. Further \ixpe observations in cooperation with radio ones are fundamental to correlate the jet emission direction with the PA, studying the possible impact of the relativistic precession of the inner disk and/or the possible variation of the corona geometry with the different accretion states.

\section*{Acknowledgments}
This research used data products provided by the IXPE Team (MSFC, SSDC, INAF, and INFN) and distributed with additional software tools by the High-Energy Astrophysics Science Archive Research Center (HEASARC), at NASA Goddard Space Flight Center (GSFC).  IXPE is a joint US and Italian mission. FLM is supported by the Italian Space Agency (Agenzia Spaziale Italiana, ASI) through contract ASI-INAF-2022-19-HH.0,  by the Istituto Nazionale di Astrofisica (INAF) in Italy, and partially by MAECI with grant CN24GR08 “GRBAXP: Guangxi-Rome Bilateral Agreement for X-ray Polarimetry in Astrophysics”.
\bibliographystyle{yahapj}
\bibliography{bib}

\end{document}